\documentclass[aps,prl,preprint,showpacs,groupedaddress]{revtex4}
\usepackage {amssymb}
\usepackage {amsmath}
\usepackage {graphicx}
\usepackage {longtable}

\begin{document}
\title{The high-temperature phase transition in a plasma}

\author{Fedor V.Prigara}
\affiliation{Institute of Microelectronics and Informatics,
Russian Academy of Sciences,\\ 21 Universitetskaya, Yaroslavl
150007, Russia} \email{fprigara@imras.yar.ru}

\date{\today}

\begin{abstract}

The line of phase equilibrium between the hot and cold phases of a
plasma is derived from the intensity of thermal radiation as a
function of the plasma density and temperature. The last function
is obtained with the help of the condition for emission relating
the wavelength of radiation and the plasma density.

\end{abstract}


\maketitle

It was shown recently [1] that a transport barrier in toroidal plasma, i.e.
the region of reduced energy transport across magnetic surfaces [2-4], is
the boundary region between the hot and cold phases of the plasma. The
transport properties of the hot phase essentially differ from those of the
cold phase due to the intense interaction of the plasma with thermal
radiation above the inversion temperature [5]. The boundary region between
the hot and cold phases corresponds to the minimum values of specific
resistivity and transport coefficients of the plasma. Other specific
properties of a transport barrier such as low magnetic shear, the generation
of radial electric fields, and the poloidal rotation are the consequences of
the minimum resistivity and transport coefficients of a plasma in this
region.

The specific resistivity of the cold phase in the vicinity of the phase
transition to the hot phase is very small, so the phase transition is nearly
superconductive. The additional current flows in the cold phase nearby the
boundary with the hot phase and does not heat the hot phase, so the
additional neutral-beam or radio-frequency heating is required to obtain
higher temperatures in the hot phase [3,4].

The critical value $T_{e,c} $ of the electron temperature,
corresponding to the transition from the cold to the hot phase,
for relatively low plasma densities achievable in tokamaks is
independent of the density and has an order of magnitude of the
inversion temperature $T_{0} \cong 2keV$ [5]. Studies of sawtooth
oscillations by von Goeler, Stodiek, and Southoff [6] suggest that
the transition temperature is approximately $T_{e,c} \cong
0.8keV$. In this paper, we expand the line of phase equilibrium
between the hot and cold phases of a plasma in the region of
higher densities where the dependence of the transition
temperature on the plasma density is essential.

Earlier, the energy distribution of atoms in the field of thermal
blackbody radiation was obtained [7] in the form

\begin{equation}
\label{eq1}
N/N_{0} = \sigma _{a} \omega ^{2}/\left( {2\pi c^{2}}
\right)\left( {exp\left( {\hbar \omega /kT} \right) - 1} \right),
\end{equation}

\noindent where $N_{0} $ is the population of the ground state
$E_{0} $, \textit{N} is the population of the energy level $E =
E_{0} + \hbar \omega $, $\sigma _{a} $ is the absorption
cross-section, $\hbar $ is the Planck constant, and \textit{T} is
the radiation temperature.

This distribution is valid in the range $\hbar \omega /kT
\geqslant 1$, since in the limit $\hbar \omega /kT \to 0$ the line
width is going to infinity, that indicates the violation of the
one-particle approximation used in [7].

The function (\ref{eq1}) has a maximum at $\hbar \omega _{m} =
1.6kT$. When the temperature exceeds the critical value of $T_{0}
= 2 \times 10^{7^{}}K$ (the inversion temperature), the population
of the energy level \textit{E} exceeds the population of the
ground state $E_{0} $. Since the function (\ref{eq1}) is
increasing in the range $\omega < \omega _{m} $ , the inversion of
the energy level population is produced also in some vicinity of
$\omega _{m} $ (below $\omega _{m} $). This suggests the maser
amplification of thermal radio emission in continuum by a hot
plasma with the temperature exceeding the critical value $T_{0} $.
Maser amplification in compact radio sources was assumed earlier
in [8] based on the high brightness temperatures of active
galactic nuclei. Since a hot plasma in an accretion disk is
concentrated nearby the central energy source, maser amplification
is characteristic for compact radio sources.

It is clear that, when the temperature of a plasma is below $T_{0}
$, the radio flux is very small, and when the temperature exceeds
$T_{0} $, radio emission is on. This an on-off cycle is detected
in the radio pulsar PSR B1259-63 [9]. Similar is an on-off cycle
in X-ray pulsars, e.g., the 35-day cycle in Her X-1. It implies
that X-ray emission from X-ray pulsars is produced by the laser
amplification in continuum which is quite analogues to maser
amplification at radio wavelengths.

It was shown also [7] that thermal emission has a stimulated
character. According to this conception thermal emission from
non-uniform gas is produced by an ensemble of individual emitters.
Each of these emitters is an elementary resonator the size of
which has an order of magnitude of mean free path \textit{l} of
photons

\begin{equation}
\label{eq2} l = \frac{{1}}{{n\sigma} }
\end{equation}

\noindent where \textit{n} is the number density of particles and
$\sigma $ is the absorption cross-section.

The emission of each elementary resonator is coherent, with the
wavelength

\begin{equation}
\label{eq3} \lambda = l,
\end{equation}

\noindent and thermal emission of gaseous layer is incoherent sum
of radiation produced by individual emitters.

An elementary resonator emits in the direction opposite to the
direction of the density gradient. The wall of the resonator
corresponding to the lower density is half-transparent due to the
decrease of absorption with the decreasing gas density.

The condition (\ref{eq3}) implies that the radiation with the
wavelength $\lambda $ is produced by the gaseous layer with the
definite number density of particles \textit{n} .

The condition (\ref{eq3}) is consistent with the experimental
results by Looney and Brown on the excitation of plasma waves by
electron beam [10].

The intensity of thermal blackbody radiation is given by the Planck formula

\begin{equation}
\label{eq4}
B_{\nu}  \left( {T} \right) = 2h\nu ^{3}/c^{2}\left(
{exp\left( {h\nu /T} \right) - 1} \right)^{ - 1},
\end{equation}

\noindent where $\nu = c/\lambda $ is the frequency of radiation,
\textit{c} is the speed of light, \textit{h} is the Planck
constant, \textit{T} is the temperature, the Boltzmann constant
being assumed to be included in the definition of the temperature.
Using the condition for emission (\ref{eq3}), we obtain the
intensity of thermal blackbody radiation as a function of the
density, \textit{n}, and the temperature, \textit{T}, in the form

\begin{equation}
\label{eq5} B\left( {n,T} \right) = \left( {2T^{3}/h^{2}c^{2}}
\right)\left( {\alpha n/T} \right)^{3}\left( {exp\left( {\alpha
n/T} \right) - 1} \right)^{ - 1},
\end{equation}

\noindent
where $\alpha = \frac{{1}}{{4}}hc\sigma _{a} $, and$\sigma _{a} $ is the
absorption cross-section.

Since the function $f\left( {x} \right) = x^{3}/\left( {e^{x} - 1} \right)$
has a maximum at $x_{m} \approx 3$, the value of the density, $n_{c} $,
corresponding to the maximum value of the intensity of thermal radiation, is
given by the formula

\begin{equation}
\label{eq6} n_{c} \approx 3T/\alpha = 12T/\left( {hc\sigma _{a}}
\right).
\end{equation}

A hot plasma with the temperature $T_{e} \geqslant T_{0} \cong
2keV$ is intensely interacting with the field of thermal
radiation. At temperatures $T \geqslant T_{0} $ the stimulated
radiation processes dominate this interaction [5]. Thermal
radiation induces radiative transitions in the system of electron
and ion which corresponds to the transition of electron from the
free to the bounded state.

Thus, in a hot plasma interacting with thermal radiation, the
bounded states of electrons and ions restore, leading to the
change of collision properties of a hot plasma. In this case, the
electron-ion collision cross-section has an order of magnitude of
the atomic cross-section, $\sigma _{0} \cong 10^{ - 15}cm^{2}$.

In the Wien region $n > n_{c} \left( {T} \right)$ the intensity of thermal
radiation exponentially decreases with the increase of the density, so the
interaction between a plasma and thermal radiation is weak, and hence this
region corresponds to the cold phase of a plasma. The hot phase of a plasma
is realized in the Rayleigh-Jeans region $n < n_{c} \left( {T} \right)$
(assuming that the temperature exceeds the inversion temperature), where the
intensity of thermal radiation comparatively weakly depends on the plasma
density.

As a result, the line of phase equilibrium between the hot and cold phases
of a plasma is described by the following approximate expressions:

\begin{equation}
\label{eq7}
 T_{c} \cong T_{c,0}, n < n_{c,0},
\end{equation}

\begin{equation}
\label{eq8}
 n_{c} \cong 12T/\left(
{hc\sigma _{a}}  \right), n > n_{c,0},
\end{equation}

\noindent where $n_{c,0} \cong 0.8 \times 10^{17}cm^{ - 3}$ for
$T_{c,0} \cong 0.8keV$, according to the equation (6).

Consider now the phase equilibrium between the hot and cold phases of a
plasma. The pressure in the cold phase is given by the formula

\begin{equation}
\label{eq9} P_{c} = 2n_{i} T,
\end{equation}

\noindent
where $n_{i} = n_{e} $ are the ion and electron densities respectively. The
pressure in the hot phase is

\begin{equation}
\label{eq10} P_{h} = n_{0} T + 2\left( {n_{h} - n_{0}}  \right)T,
\end{equation}

\noindent
where $n_{0} $ is the density of neutral atoms, and $n_{h} $ is the total
density of ions and neutral atoms.

If $n_{0} \approx n_{h} $, then from the condition of equilibrium, $P_{c} =
P_{h} $, we find that $n_{h} \approx n_{c} $, i.e. the density of the hot
phase is approximately twice of the density of the cold phase. It means that
there is the discontinuity of density at the boundary between the hot and
cold phases, analogous to the first order phase transitions.

There is also the discontinuity of resistivity somewhat analogous to the
superconduction phase transitions. The specific resistivity of the hot
phase, $\eta _{h} $, is about three orders of magnitude larger than that of
the cold phase, $\eta _{c} $. As much as large is the discontinuity of
thermal conductivity across magnetic surfaces, so the cold phase is a good
thermal protector for the hot phase.

The discontinuity of diffusion coefficients with respect to the diffusion
across magnetic surfaces leads to the generation of radial electric fields
in the region of a transport barrier due to the differences in diffusion
coefficients for ions and electrons . The drift of electrons and ions in
these electric fields produces the poloidal rotation of the plasma.

To summerise, the critical temperature of the phase transition
between the cold and hot phases of a plasma is independent of the
plasma density for low densities and increases proportionally to
the plasma density for higher densities exceeding some critical
value. There are discontinuities of density, resistivity, and
transport coefficients at the point of phase transition. The
high-temperature phase transition in a plasma is experimentally
observed as a transport barrier in toroidal plasma.

\begin{center}
---------------------------------------------------------------
\end{center}

[1] F.V.Prigara, Europhys. Lett. (submitted), E-print archives,
physics/0503197.

[2] F.Wagner et al. Phys. Rev. Lett. \textbf{49}, 1408 (1982).

[3] K.A.Razumova, Usp. Fiz. Nauk \textbf{171}, 329 (2001)
[Physics-Uspekhi \textbf{44}, 311 (2001)].

[4] A.C.C.Sips, Plasma Phys. Control. Fusion \textbf{47}, A19
(2005).

[5] F.V.Prigara, Phys. Plasmas (submitted), E-print archives,
physics/0404087.

[6] S.von Goeler, W.Stodiek, and N.Sauthoff, Phys. Rev. Lett.
\textbf{33}, 1201 (1974).

[7] F.V.Prigara, Int. J. Mod. Phys. D (submitted), E-prints
archives, astro-ph/0311532 (2003).

[8] F.V.Prigara, Astron. Nachr.,\textbf{324}, No. S1, 425 (2003).

[9] G.J.Qiao, X.Q.Xue, R.X.Xu, H.G.Wang, and B.W.Xiao, Astron.
Astrophys., \textbf{407}, L25 (2003).

[10] F.V.Prigara, Plasma Phys. Control. Fusion (submitted),
E-print archives, physics/0411207.

\end{document}